\documentstyle[psfig,epsfig,graphicx]{mn2e}

\newcommand{\be}{\begin{equation}}
\newcommand{\ee}{\end{equation}}
\newcommand{\ba}{\begin{eqnarray}}
\newcommand{\ea}{\end{eqnarray}}
\newcommand{\nn}{\nonumber \\}

\newcommand{\thetab}{\mbox{\boldmath $\theta$}}

\newcommand{\bell}{{\mbox{\boldmath{$\ell$}}}}

\newcommand{\mnras}{MNRAS}
\newcommand{\prd}{Phys. Rev. D}
\newcommand{\apjl}{Astro. Phys. Journal Letters}
\newcommand{\apj}{Astro. Phys. Journal}
\newcommand{\aap}{AAP}

\def\gs{\mathrel{\raise1.16pt\hbox{$>$}\kern-7.0pt %
\lower3.06pt\hbox{{$\scriptstyle \sim$}}}}         %
\def\ls{\mathrel{\raise1.16pt\hbox{$<$}\kern-7.0pt %
\lower3.06pt\hbox{{$\scriptstyle \sim$}}}}         %

\title[3D Photometric Cosmic Shear]{
3D Photometric Cosmic Shear}
\author[Kitching, Heavens \& Miller] {T. D. Kitching$^{1}$\thanks{tdk@roe.ac.uk}, A. F. Heavens$^{1}$ \& L. Miller$^{2}$\\
$^{1}$SUPA, Institute for Astronomy, School of Physics, University of
Edinburgh, Royal Observatory, Blackford Hill, Edinburgh, EH9 3HJ,
U.K.\\
$^{2}$Department of Physics, Oxford University, Keble Road, Oxford, OX1 3RH, U.K.}

\date{}

\pagerange{\pageref{firstpage}--\pageref{lastpage}}

\pubyear{2010}

\begin{document}

\maketitle

\label{firstpage}


\begin{abstract}

\noindent Here we present a number
of improvements to 
weak lensing 3D power spectrum analysis, 3D cosmic shear, that 
uses the shape and redshift information of every
galaxy to constrain cosmological parameters.
We show how 
photometric redshift probability distributions for individual
galaxies can be directly included in this statistic with no averaging. 
We also include the Limber approximation, considerably simplifying full
3D cosmic shear analysis, and we investigate its range of applicability. 
Finally we show the relationship between weak lensing tomography and
the 3D cosmic shear field itself; the steps connecting them being the Limber
approximation, a harmonic-space transform and a discretisation in wavenumber. 
Each method has its advantages: 3D cosmic shear analysis allows
straightforward inclusion of all relevant modes, thus  
ensuring minimum error bars, and direct control of the range of physical 
wavenumbers probed,  to avoid the uncertain highly nonlinear regime.
On the other hand, tomography is more convenient for  
checking systematics through direct investigation of the redshift
dependence of the signal.  Finally, for tomography, we suggest that
the angular modes  
probed should be redshift-dependent, to recover some of the 3D advantages.

\end{abstract}


\section{Introduction}

Cosmology is faced with a standard model that contains two dominant
unknown components (dark energy and dark matter) and two untested
assumptions (that general relativity applies on large scales and that
there was an inflationary period in the early Universe). In order to
move forward from this unsatisfactory situation the need for high
precision methods, which exhibit a strong statistical signal to
cosmological parameters (those that describe the standard model or 
deviations from it) to allow various models to be compared with a high
discriminatory power, is obvious. 

It has become clear that 3D cosmic shear (Heavens, 2003; Castro,
Heavens, Kitching, 2005; Heavens et al., 2006; Kitching et al., 2007)
and weak lensing tomography 
(Hu 1999; Amara \& Refregier, 2007; Bernstein \& Jain, 2006),
in which galaxy 
redshifts and the weak lensing shear distortion are used
simultaneously in the signal, is an approach which is
remarkably sensitive to cosmological parameters through both the
growth of structure and the geometric behaviour of the lensing effect
itself. For example neutrino mass (Hannestaad et al., 2007; 
Kitching, Heavens, Verde, 2007; de Bernardis
et al., 2009; Jimenez et al., 2010), 
modified gravity (Thomas et al., 2008; 
Heavens, Kitching, Verde, 2008),
dark matter (Camera et al., 2010; see Massey et al., 2010 for a 
recent review). 
In particular, given a current
focus on dark energy 
(Albrecht et al., 2006; Peacock \& Schneider, 2006) 
as the point in which we may find a departure from our standard model,
it has been shown (and is widely accepted) that 3D cosmic shear
has the potential to place tight constraints on the equation of state
of dark energy. 

This article is concerned with the techniques outlined in 
Heavens (2003), Heavens et al. (2006) and Kitching et al. (2007) 
whereby the 3D shear field (a function of redshift and angle) is
treated in its entirety. In this approach the 3D shear power spectrum is
reconstructed using spherical harmonics and the cosmological signal is found
in the covariance of the coefficients. This is an alternative to weak
lensing tomography where the 3D shear field is binned and projected in
redshift space to produce a series of 2D projections, from which the
auto- and cross-power spectra can be calculated 
(e.g. Hu 1999; Amara \& Refregier, 2007, Schrabback et al., 2009).

3D cosmic shear has some advantage over tomography in that no binning is required, avoiding
its consequent loss of information.  One of the purposes of this
article is to explore the relationship  
between 3D cosmic shear and tomography.
3D cosmic shear in principle allows a full sky reconstruction
including the effects of 
sky curvature.  Individual pairs of galaxies in redshift
can be excluded from the estimator directly (to reduce intrinsic
alignment contamination; King \& Schneider, 2002l; 
Heymans \& Heavens, 2003) and
since every galaxy contributes directly and individually to the signal
information such as the probability distribution in redshift of
individual galaxies can be included in the estimator. 

This final
point, the inclusion of individual redshift probabilities, has been
alluded to but has not been implemented up until now in 3D
cosmic shear. Previously the
inclusion of photometric redshift uncertainties has been through the
use of the average photometric redshift error distribution at each redshift. We
will show that this approach can lead to a mis-estimate of cosmological errors, and
introduces small biases in cosmological parameter. 

We will also present a Limber approximation of the full 3D shear field
and show that for the range of (radial and azimuthal) modes which 
are sensitive to
cosmological information the approximation is very good. This
approximation vastly simplifies the theoretical computation of the 3D
covariance, making it as simple as tomography to apply. We also
explicitly derive the weak lensing tomography power spectra from the
full 3D shear field, showing that these are related by two 
approximations (the Limber approximation, and the binning
approximation), and demonstrate that tomography provides a sampling of
the physical wavenumber space which is $\ell$- and bin-dependent. 

This paper is organised as follows. In Section \ref{Photometric 3D
  Shear Estimator} we will review the 3D cosmic shear formalism and
show how photometric redshift information from individual galaxies can be
explicitly included, using Fisher matrices we will investigate the
impact on parameter error and bias. In Sections \ref{Limber
  Approximation} and \ref{Tomography from 3D Cosmic Shear} 
we will present 
several approximations of the 3D shear field, using the Limber
approximation. We discuss results in Section \ref{Conclusion}. 

\section{Photometric 3D Shear Estimator}
\label{Photometric 3D Shear Estimator}

3D cosmic shear decomposes the 3D shear field (a function of angular and
radial position) using spherical harmonics (see Heavens, 2003; Castro,
Heavens, Kitching, 2006; Heavens et al., 2006). Using the flat
sky approximation we replace spherical harmonics with exponential functions, and assuming the Universe is flat 
we replace the more general ultra-spherical Bessel functions by spherical Bessel functions, 
a 3D shear estimator can be written, by summing over galaxies
\be 
\label{1}
\hat\gamma_i(k,\bell)=\sqrt{\frac{2}{\pi}}\sum_g\gamma_i^g
j_{\ell}(k r^g_0){\rm e}^{-i\bell.\thetab^g}W(r^g_0)
\ee
where $\gamma_i^g$ is the $i^{\rm th}$ shear component for each galaxy
$g$ at angular position $\thetab$ and radial position $r^g_0$, 
which is obtained from the photometric redshift assuming a fiducial cosmology. $W(r)$ is an
optional weight which we set to $W=1$ for clarity in the remainder of
this article, we
note here that there may exist optimal weights, particularly for
constraining a certain cosmological parameter set. $r^g_0$ is a
distance, not a redshift, so to convert from some data
(spectra/photometry $\rightarrow$ redshifts) requires the assumption of a 
cosmology -- however as shown in Kitching et al. (2007) this
assumption is benign. Note that the transform convention of Castro et al. (2005)
includes an extra factor of $k$, which we omit. The result of this expansion is
a set of four 3D data vectors that are functions of angular
wavenumber ($\bell=2\pi/\btheta$) and radial wavenumber and  
a real and imaginary part of $\gamma_1$ and $\gamma_2$.   Note that
the behaviour of the Bessel functions (that $j_{\ell}(x)\ll 1$ for
  $x \ll \ell$)  ensures that a mode of radial
wavenumber $k$ also has no transverse power with smaller wavelengths, so we are justified
in identifying $k$ as a physical wavenumber in 3D space. 
The assumption of a flat sky (resulting in an exponential in place of $Y^m_{\ell}$ above) 
can be relaxed (see Castro et al., 2005), resulting in 
a covariance of this estimator that is similar to the one that we derive below, 
we leave a full investigation of this for a future article. The assumption of 
a flat Universe (resulting in Bessel functions in place of ultra-spherical Bessel functions) 
can also be relaxed, for computational and notational convenience we use Bessel functions 
in this article and take small perturbations about this model.

Since the mean of the estimators is zero we use the covariance
(Tegmark, Taylor \& Heavens, 1997)
of the estimator to constrain cosmology. In Appendix A we derive an
expression for the expected covariance of the 3D shear estimators
keeping track of any sum over galaxies.  We may write  the components of the
shear estimator by 
$\hat\gamma_1(k,\bell)=\frac{1}{2}\left(\hat\Gamma_{11}-\hat\Gamma_{22}\right)$
and $\hat\gamma_2(k,\bell)=\hat\Gamma_{12}=\hat\Gamma_{21}$, where the
$\Gamma$ terms are defined in terms of the lensing potential  
$\phi$ by $\Gamma_{ij}=\tilde\partial_i\tilde\partial_j\phi$, and have 
covariance
\ba 
\label{eq2}
&&C^{3D}_{ij,\ell}(k_1,k_2)=\langle \hat\Gamma_{ij}(k_1,\bell)
\hat\Gamma^*_{ij}(k_2,\bell)\rangle=\nn&&
\Delta\Omega \ell^2_i\ell^2_j 
\left(\frac{2}{\pi}\right)^2\left(\frac{4}{c^2}\right)A^2
\sum_g\sum_h j_{\ell}(k_1 r^g_0)j_{\ell}(k_2 r^h_0)\nn
&& \int_0^{r^g_0}{\rm d}r'\int_0^{r^h_0}{\rm d}r''
F_K(r^g,r')F_K(r^h,r'')\nn
&&\int \frac{{\rm
    d}k'}{k'^2}\frac{1}{a(r')a(r'')} 
j_{\ell}(k'r')j_{\ell}(k'r'')\sqrt{P(k';r')P(k';r'')}.
\ea
$F_k(r,r')=S_k(r-r')/[S_k(r)S_k(r')]$ ($S_k=\sinh(k),k,\sin(k)$
for open, flat or closed geometries so $F_k(r,r')=(r-r')/rr'$ for a
flat universe), $a(r)=1/[1+z(r)]$ is the scale factor and $P(k; r)$ is
the 3D matter power spectrum. 
The prefactors to this expression are
$\Delta\Omega$ the dimensionless area of the survey, 
$A=3\Omega_mH^2/2$ where $\Omega_m$ is the present dimensionless matter
density and $H$ is the present Hubble expansion rate. The
$\ell^2_i\ell^2_j$ is a factor that is introduced
  as a result of the derivative taken when 
transforming the lensing potential to the shear (see Heavens et
al., 2006 and Kitching et al., 2007). 
The $\ell$ modes are uncorrelated due to the
assumption of isotropy.
To fully describe a non-Gaussian random field higher order potential
correlations are required (see Munshi et al., 2010).
The $r^g_0$ and $r^h_0$ are calculated from 
photometric redshifts assuming a fiducial model, where the $r^g$ and $r^h$
are their true values. This expression is
one of the main results of this 
article. 
\begin{figure*}
 \includegraphics[angle=0,width=0.95\columnwidth]{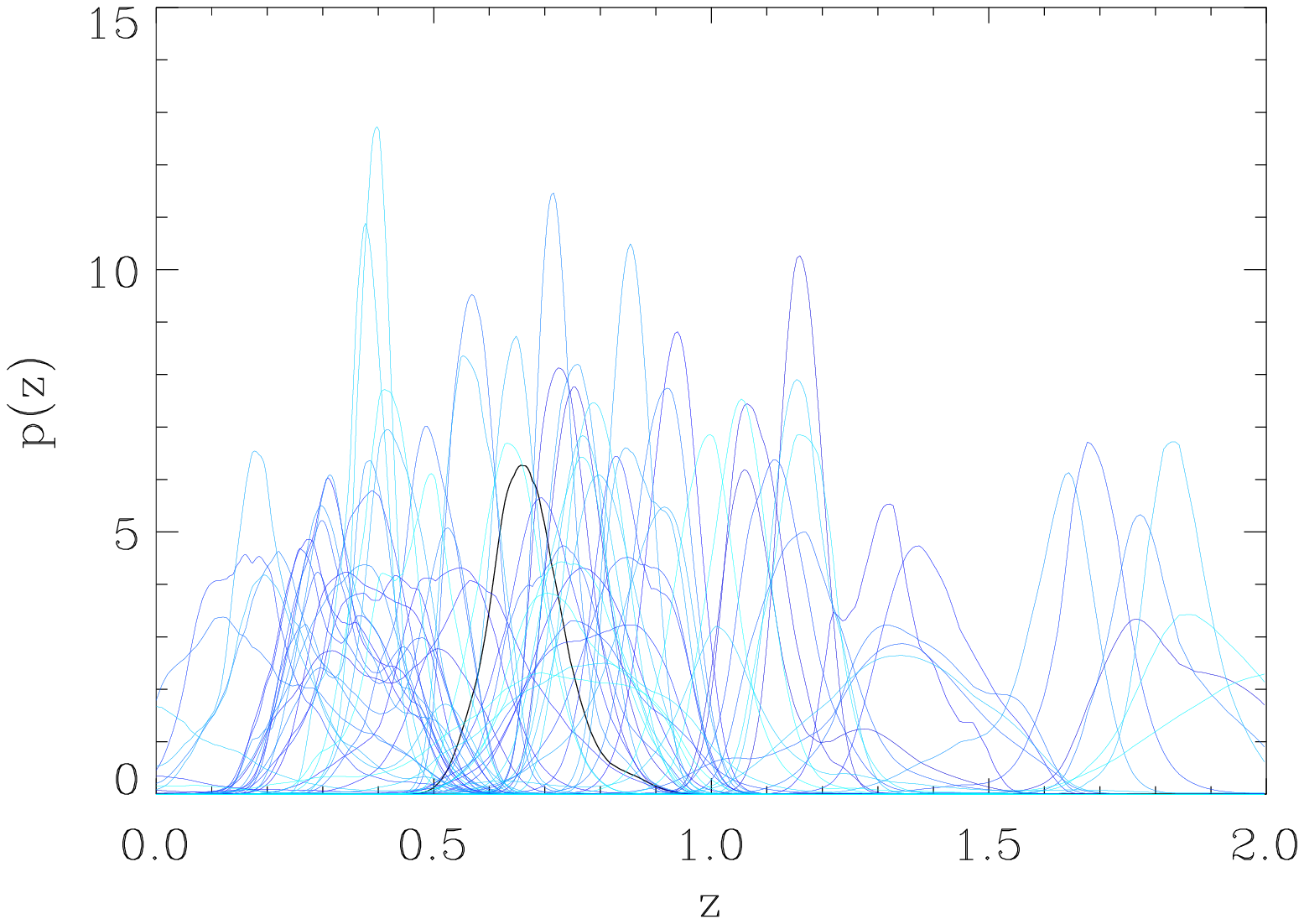}
 \includegraphics[angle=0,width=0.95\columnwidth]{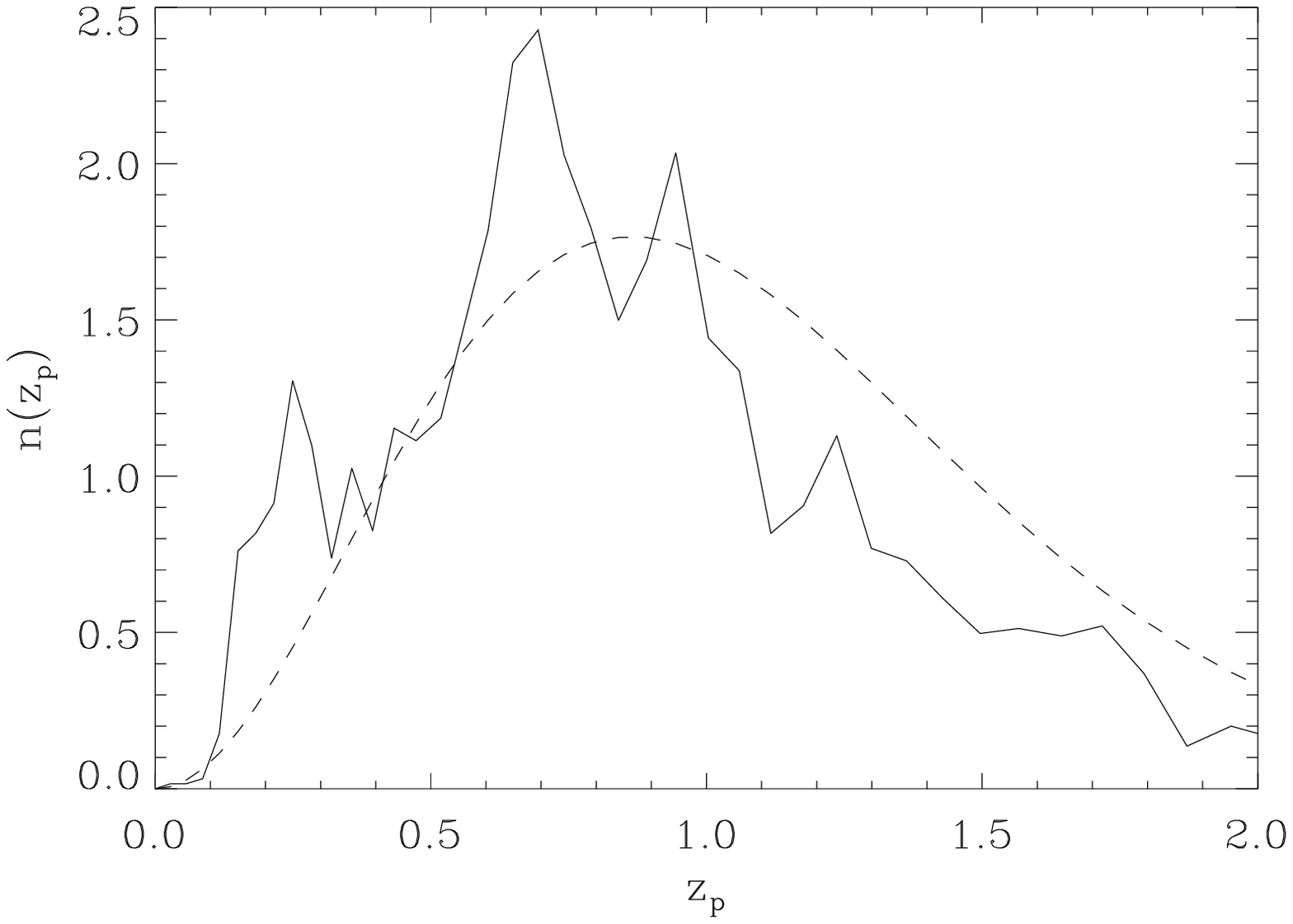}
 \caption{Left panel: a sample of $50$ redshift
   distributions from the sample used (from Bordoloi et al., 2009). 
   Right panel: the
   $n(z_p)=\sum_g p(z_p)$ from the total set of $p(z)$,
   the dashed line shows the functional form for $n(z)\propto
   z^2\exp[-(1.4z/z_m)^{1.5}]$ from Taylor et al., (2007) with
   $z_m=1.0$ for comparison.}
 \label{realpz}
\end{figure*}

In making any parameter error forecasts we will use the Fisher matrix
formalism (Tegmark, Taylor, Heavens, 1997) and assume Gaussian
likelihood surfaces, and Gaussian 
distributions for the data. For 3D cosmic shear the Fisher matrix was
introduced in Heavens (2003) and examined in more detail in
Heavens et al., (2006) and Kitching et al., (2007). 
Since we are assuming that the parameters affect 
the covariance, not the mean, the Fisher matrix is given by 
\be 
\label{fish}
F_{\alpha\beta}=\frac{g}{2}\int{\rm d}\phi_{\ell}\int{\rm d}\ell\ell {\rm
  Tr}[C^{-1}_{\ell}C_{\ell},_{\alpha}C^{-1}_{\ell}C_{\ell},_{\beta}]
\ee
where we include an integral over $\ell$-space\footnote{The density of
  states accounts for correlations between modes arising from partial
  sky coverage, equivalent to the $f_{sky}$ approach of many papers.
  Note that the insensitivity to large-scale modes, which is also a
  consequence of using a patch of sky,  needs to be treated by a cut
  on $\ell$.  The Fisher matrix approach assumes the data are
  Gaussian; see  Munshi et al., (2010) for an investigation of
  non-Gaussianity in 3D shear.}  
which includes a density of states in $\ell$-space, $g=$Area/$(2\pi)^2$ 
(see Appendix B of Kitching et al.,
2007).  A comma represents a derivative with respect to parameter
$\alpha$ or $\beta$. The covariance $C$ is a contribution of signal terms
and (shot) noise terms $C=C^{\rm signal}+C^{\rm noise}$, where the
signal is defined in equation (\ref{eq2}) (we have suppressed the
notation $C^{3D}_{ij,\ell}(k_1,k_2)\rightarrow C_{\ell}$ for clarity in equation 
\ref{fish}). The shot
noise covariance matrix used is the same as that in Heavens et al.,
(2006). 

Throughout we will use the parameter set (with fiducial values) :
$\Omega_m (0.3)$, $\Omega_{de}(0.7)$, $w_0(-0.95)$, $w_a(0.0)$, $h(0.71)$,
$\Omega_b(0.045)$, $\sigma_8(0.8)$ and $n_s(1.0)$. We use the
expansion of the dark energy equation of state as introduced in
Chevallier \& Polarski (2001) and Linder
(2003), we include generally curved geometries where
$\Omega_k=1-\Omega_m-\Omega_{de}$\footnote{In this case 
ultra spherical Bessel functions should be used, but as pointed out in
Kitching et al. 2007 in the $\ell\gg 1$ and $k\gg ({\rm
  curvature})^{-1}$ regime the use of the $j_{\ell}(kr)$ is well justified.}
. Our matter power spectrum is
constructed using 
the Eisenstein \& Hu (1999) transfer functions, with baryon wiggles,
and employ 
the Smith et al. (2003) non-linear correction; for curved cosmologies with varying
$w(z)$ we use the {\tt iCosmo} (Refregier et al., 2008) interpolation scheme
(also discussed in Schrabback et al., 2009). 

Throughout we do not include any systematics effects in the predicted
parameters errors. However we note that the impact from all primary
systematics (photometric redshift calibration, intrinsic alignments
and shape measurement error) should be at most a factor of $\sqrt 2$
on cosmological parameters (Kitching et al., 2008) in a realistic
self-calibration regime using parameterised systematic
  functions. Using non-parametric approaches may degrade results further, but
  as shown in Kitching \& Taylor (2010) and Kitching et al. (2009),
  using tomographic analyses, the overall cosmological impact is not severe. 

We assume a fiducial survey configuration of $20$,$000$ square
degrees, with a surface number density of $35$ galaxies per square
arcminute and an intrinsic ellipticity dispersion of $0.3$. Where we
do not use direct photometric redshift probabilities we will use a
redshift distribution of galaxies $n(z)\propto z^2{\rm
  exp}[-(1.4z/z_m)^{1.5}]$ given in Taylor et al.,
(2007) with a median redshift of $z_m=1$ and a Gaussian redshift dispersion
with a redshift error $\sigma_z(z)=0.03(1+z)$. This survey configuration
is similar to the proposed Euclid wide survey (Refregier et al.,
2010). Throughout we use a maximum azimuthal wavenumber of $\ell_{\rm
  max}=5000$ and a maximum radial wavenumber of $k_{\rm
  max}=1.5$hMpc$^{-1}$ to avoid the highly non-linear regime where
theoretical predictions for the power spectrum may be unsound (e.g. 
Rudd et al., 2008; Joudaki et al., 2009).

\subsection{Including Photometric Redshifts}
\label{Including Photometric Redshifts}

For the depth of surveys required for optical weak lensing, it is
impractical to obtain spectroscopic redshifts, 
so photometric redshifts $z_p$ are obtained, typically from broad-band
photometry.  Some photo-z codes return a complete posterior
probability distribution for the  
redshift, given the photometry, and the purpose of this section is to
show that this individual information can be used when it exists.
Since it uses all the information available, it should yield a more
accurate covariance matrix than the alternative of reducing the
redshift distributions 
to a simpler $\bar p(z|z_p)$; the form of $p$ may be arbitrary
(and include for example outliers), but its distribution is assumed in
the alternative approach to be the same for all galaxies with the same
photometric redshift. Note that, although a full posterior redshift
distribution may be returned, we still do have to identify a unique
photo-z for each galaxy, to provide an $r_0^g$ to insert into the
spherical Bessel function weighting. 

To include galaxy photometric redshift errors we integrate over the
posterior redshift distribution for each galaxy.  We write this as
$p_g(z|z_g)$ as a shorthand for the redshift probability distribution
given all the photometric information available for galaxy $g$:  
\ba 
\label{cov}
&&C^{3D}_{ij,\ell}(k_1,k_2)=
\Delta\Omega\ell^2_i\ell^2_j\frac{4}{\pi^2c^2}A^2
\sum_g\sum_h
\left[j_{\ell}(k_1 r^g_0)j_{\ell}(k_2 r^h_0)\right]\nn
&&\int{\rm d}z' p_g(z'|z_g) \int{\rm d}z'' p_h(z''|z_h) \nn
&&\int_0^{r'}{\rm d}{\tilde r}'\int_0^{r''}{\rm d}{\tilde r}''
F_K(r',{\tilde r}')F_K(r'',{\tilde r}'')\nn
&&\int \frac{{\rm
    d}k'}{k'^2}\frac{1}{a({\tilde r}')a({\tilde r}'')} 
j_{\ell}(k'{\tilde r}')j_{\ell}(k'{\tilde r}'')\sqrt{P(k';{\tilde r}')P(k';{\tilde r}'')},
\ea
where $r'=r(z')$. 
This is the covariance of the full 3D shear field constructed only 
from a sum over the individual tracer galaxy population, and including
all uncertainty that may be present from photometric redshift
estimates. 

In this formalism we can
retain individual posteriors.  Since we do not want the data vector
to depend on the cosmological parameters (and consequently violating
the conditions of the Fisher matrix analysis), we must assume a
fiducial cosmology to translate a photo-z to a distance $r_0^g$. 
However as long as
this choice, for each galaxy, is the same as those used in the actual
sum over data (equation \ref{1}) then the estimator will be
unbiased.   

For clarity we can re-express the covariance as series of matrices
\ba
\label{CGU}
C^{3D}_{ij,\ell}(k_1,k_2)
&=&{\mathcal A}^2\int \frac{{\rm d}k'}{k'^2}G_{\ell}^D(k_1,k')G_{\ell}^D(k_2,k')\nn
G_{\ell}^D(k_1,k')&=&\sum_g j_{\ell}(k_1r^g_0)\int {\rm
  d}z'p_g(z'|z_g) U_{\ell}(r(z'),k')\nn
&=&\int {\rm
  d}z'\left[\sum_g j_{\ell}(k_1r^g_0)p_g(z'|z_g)\right] U_{\ell}(r(z'),k')\nn
U_{\ell}(r(z'),k')&=&\int_0^{r(z')}{\rm d}{\tilde
  r}\frac{F_K(r(z'),{\tilde r})}{a({\tilde r})}j_{\ell}(k'{\tilde
  r})P^{\frac{1}{2}}(k';{\tilde r})
\ea
where ${\mathcal A}^2=\Delta\Omega\ell^2_i\ell^2_j\frac{4}{\pi^2c^2}A^2$. Equation
(\ref{CGU}) is a discrete case of the equations in Section 2.5 of
Heavens et al. (2006). The $ij$ refer to $\gamma$ combinations (see
Appendix A), not redshift bins -- this is a continuous 3D estimator. 

If we do not have individual posterior distributions, or wish to
ignore the individual galaxy redshift error 
distributions, we can simplify $p_g(z|z_g)$ to a global $\bar
p(z|z_p)$ 
so all galaxies at fixed photometric redshift are assumed to have the same
distribution of true redshifts. In this regime the 
the $G$ matrix is modified only so that 
\ba
\label{Gcont}
G^C_{\ell}(k_1,k')=\int {\rm d}z_p{\rm d}z'j_{\ell}(k_1r(z_p))
n(z_p)\bar p(z'|z_p)U_{\ell}(r(z'),k')
\ea
where $\bar p(z|z_p)$ is the redshift probability distribution
at photometric redshift $z_p$, and $n(z_p)$ 
is the number density of galaxies 
as a function of redshift. Equation (\ref{Gcont}) 
is in agreement with Castro et al. (2003) and Heavens et
al. (2006). 

By including the individual galaxies in equation (\ref{CGU}) we automatically take
into account all overlap between photometric redshift posteriors, all outliers in the
sample and have the best estimate for the covariance of the data vector. 
Of course the outliers have only been properly accounted for only if the $p(z|z_p)$ 
is a correct estimate of the probability of a galaxy being an outlier; we leave an 
investigation into the effect of errors on the $p(z|z_p)$ for future study.

\subsection{Forecasted Impact}

To illustrate the effects, 
we use a sample of realistic photometric redshift probabilities
$\{p(z_p)\}$ from Bordoloi et
al. (2009), which are a simulated set of redshift probability
distributions like those expected from Euclid. We show these in Figure
\ref{realpz}. We use a representative sample of $3000$
redshift distributions, for computational speed, which for a surface
number density of $n_0=35$ per square arcmin represents a sample of
galaxies from around $100$ square arcmin. This is low compared to the
expected number of galaxies in future surveys, however we use this
sample as representative for the investigation of the approximations
we highlighted in Section \ref{Including Photometric Redshifts} -- in
fact with low(er) number statistics we may expect a larger deviation
than in reality because the error on the mean will be larger (a $1/N$
effect). 
To match the total number to the surface density we
multiply the signal covariance by an extra factor of $(n_0/3000)^2$
(and the noise by $n_0/3000$). 

We consider two cases 
\begin{enumerate}
\item 
First case, where we use the full covariance
including posteriors for the true redshifts\footnote{In practice we 
  have used redshift-binned true photometric 
  redshift posteriors, using the resolution provided by Bordoloi et
  al. (2009), which was $5000$ bins in redshift. This is sufficient to
  illustrate the method, since this is much larger than the number of
  $p(z)$ availabel. A binned
  description is not a limitation of this method -- one could
  imagine a basis set decription with no explicit redshift binning.}, 
equations (\ref{CGU}).\\
\item 
Second case, where we make an approximation by summing the
individual posteriors in photometric redshift intervals $b$, to
find a global number density distribution as a function of redshift 
$\bar p(z|z_p)=\sum_{g:b} p_g(z|z_p)$, equation (\ref{Gcont}). This is
one particular method of smoothing the redshift distribution, we would expect
similar results with other smoothing kernels with a similar redshift
scale. In
this example we use $20$ bins in $z_p$ between the redshifts $0$--$2$
(a step in redshift of $0.1$).
\end{enumerate}

\begin{figure*}
 \centerline{
 \includegraphics[angle=0,width=1.6\columnwidth]{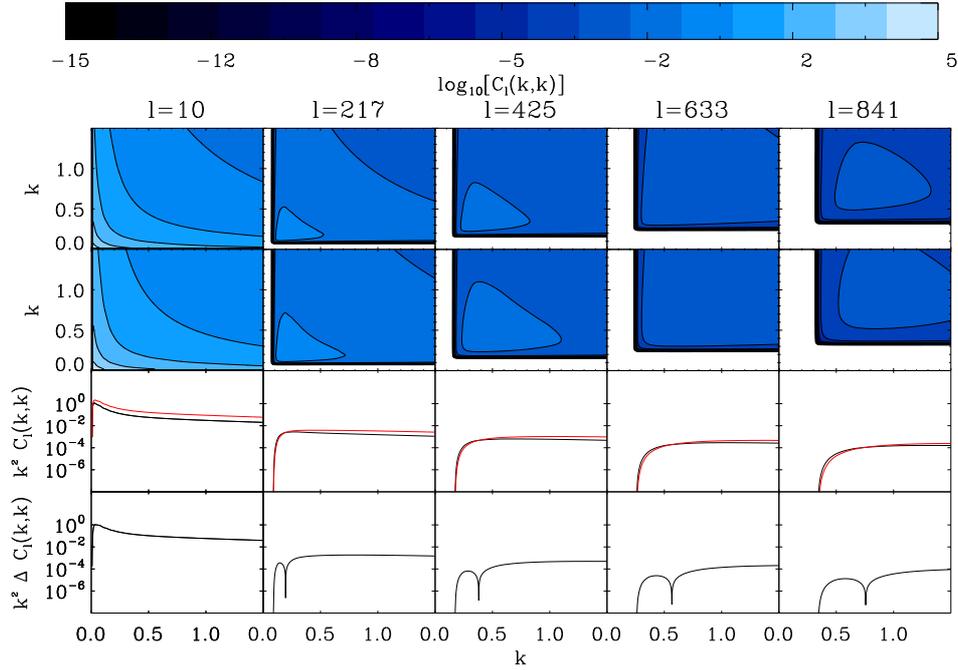}}
 \caption{The 3D cosmic shear power spectra for the  
   case where we use individual galaxy redshift distributions (upper
   panel), 
   and using an approximation of the
   photometric probability distribution (second panels). The colour scale is
   shown in the bar above. The third set of panels shows the diagonal
   $C_{\ell}(k,k)$ of the full 3D power spectra for the full 
   case (black lines) and for the approximation (red, dark
   grey lines). The bottom panels show the modulus of the  
   difference in the diagonal elements. Rightwards of the cusp in the bottom panels 
   the difference between the power spectra becomes negative. For a 
   Euclid survey with Bordoloi et al., (2009) $p(z)$.}
 \label{pzplot}
\end{figure*}

We show the $n(z)$ 
from the sample of photometric redshift posteriors we use in Figure
\ref{realpz}. In Figure \ref{pzplot} we show the impact on the 3D cosmic shear power
spectra. The power spectra are a function of $\ell$ and two physical
wavenumbers, hence we show the plane $(k_1,k_2)$ for a series of
$\ell$ modes, as well as the diagonal $C^{3D}_{\ell}(k,k)$. It can be seen
that using an approximation of the redshift distribution can cause
significant residuals in the power spectra, with residuals
of order $0.01$--$1$ over all scales. These residuals are most
prominent at low-$\ell$ and small radial scales. 
\begin{figure*}
 \centerline{
 \includegraphics[angle=0,width=1.6\columnwidth]{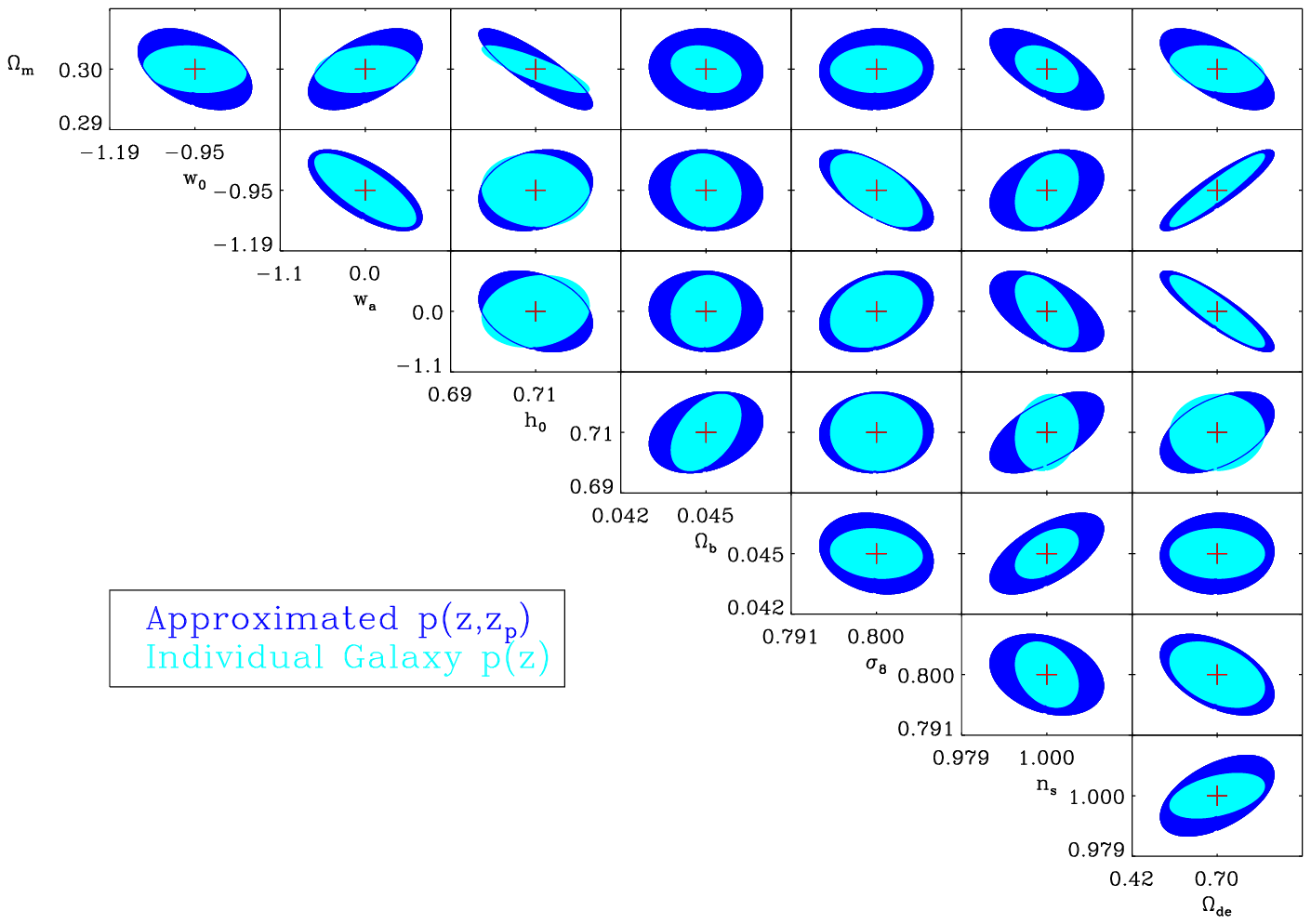}}
 \caption{The Fisher matrix $2$-parameter $1$-$\sigma$ constraints for
   the case where individual galaxy $p(z)$ are used 
   and the approximation of using a summed, approximated global
   $\bar p(z|z_p)$. 
   The central cross shows the fiducial
   values. This is for a Euclid survey (Refregier et al., 2010) using
   $p(z)$ from Bordoloi et al., (2009).}
 \label{pzfisherplot}
\end{figure*}

Using the Fisher matrix approach Figure \ref{pzfisherplot} shows 
how the approximation of averaging the posteriors $\bar p(z|z_p)$ in
photometric redshift, can affect
cosmological parameter 
errors. In general the errors are degraded by averaging the individual 
galaxy posteriors $p_g(z|z_g)$ by a factor of $10$--$50\%$, 
and in some cases the parameter degeneracies are changed. This is a
result of the individual $p_g(z|z_g)$ distributions explicitly including outlying
and non-Gaussian behaviour on a galaxy-by-galaxy basis. 

In making the approximation of $\bar p(z|z_p)$ the best fit cosmological parameter
values may also be biased, as well as the errors being affected. 
In Appendix C we show how to calculate the bias in the Fisher matrix
approximation for the case that the parameter dependency is in the
covariance (this is a generalisation of the result of Knox et al.,
1998). We find that the approximation made in this example
introduces a negligible bias on all cosmological parameters : 
$\Omega_m (-3.0\times 10^{-5})$, $\Omega_{de}(2.5\times 10^{-4})$,
$w_0(1.3\times 10^{-4})$, $w_a(-1.2\times 10^{-4})$, $h(5.2\times
10^{-4})$, 
$\Omega_b(1.1\times 10^{-6})$, $\sigma_8(7.3\times 10^{-5})$ and
$n_s(2.5\times 10^{-4})$. 
We note that changes in the covariance only impact cosmological parameter
biases if the change induced is similar to the effect of any
cosmological parameter (see for example Kitching et al. 2009; form
filling functions). In this case the changes in the covariance 
are large, but the changes are not similar to the effect of 
cosmological parameter so the biases are small. The amplitude of the 
bias will depend on the exact distribution and form of the posterior 
redshift distributions, here we use an example to introduce the methodology, 
we leave a more detailed study for a future article. 

For any given
experiment, the exact
degradation will be dependent on the exact galaxy survey and quality
of photometric redshifts available. 

Note that in Heavens et al. (2006) we
made a further approximation by assuming that the $\bar
p(z|z_p)=\bar n(z_p)p_{\rm Gauss}(z|z_p)$, where $\bar n(z_p)$ is the number
density of galaxies as a function of photometric redshift and
$p_{\rm Gauss}(z|z_p)$ is the normalised photometric redshift distribution at
each redshift which we assummed to be a Gaussian at each redshift with a 
variance $\sigma_z(z_p)$.

\section{Limber Approximation of the 3D Shear Field} 
\label{Limber Approximation} 

We now investigate various approximations of the 3D shear field. 
From LoVerde \& Afshordi (2008) the Limber approximation can be
encapsulated in the following substitution 
\be
\label{l1}
\lim_{\ell\to\infty}j_{\ell}(kr)\rightarrow \sqrt{\frac{\pi}{2\left(\ell+\frac{1}{2}\right)}}\delta^D\left(kr-\left[\ell+\frac{1}{2}\right]\right),
\ee
where $\delta^D$ is the Dirac delta function.  This is substituted
  into any integral or sum that the Bessel function appears; for example
  equations (\ref{CGU}) and (\ref{Gcont}). We note that this does not explicitly
convert a 3D wavevector into a 2D wavevector,
$k=(k_x,k_y,k_z)\rightarrow (k_x,k_y)$ (which is explicit in
  an alternative and complementary derivation of the Limber approximation;
Kaiser, 1998), but that this transformation 
is implicit when an integral over the delta function is performed. 

By substituting this approximation into the matrices 
in equation (\ref{CGU}) we can find an approximate expression of the
covariance in the high $\ell$ limit. 
In this case the $U$ matrix is modified to 
\be 
\label{l2}
U^{\rm Limber}_{\ell}(r(z'),k')=\sqrt{\frac{\pi}{2\left(\ell+\frac{1}{2}\right)}}\frac{F_K(r(z'),r_{\nu})}{a(r_{\nu})}P^{\frac{1}{2}}(k';r_{\nu})
\ee
where $r_{\nu}\equiv (\ell+\frac{1}{2})/k'<r(z')$.

In the individual redshift error case the $G$ matrix is modified to 
\ba 
\label{l3}
&&G_{\ell}^D(k_1,k')=\sqrt{\frac{\pi}{2\left(\ell+\frac{1}{2}\right)}}\nn
&&\sum_{\{g :
  r^g_0=[\ell+\frac{1}{2}]/k_1\}}\int {\rm
  d}z'p_g(z'|z^g) U^{\rm Limber}_{\ell}(r(z'),k')
\ea
where each galaxy 
only contributes to a given combination of $k_1$ and $\ell$ set by its
comoving distance $r^g_0=[\ell+\frac{1}{2}]/k_1$. We note that 
this equation in practice would require some binning in $k$ and $\ell$ 
to obtain a finite number of galaxies in this sum.

If a global error distribution is used, the $G$ matrix is modified to 
\ba
\label{l4}
&&G^{C, \rm Limber}_{\ell}(k_1,k')=\frac{1}{4\pi} n(z_{\nu})\nn
&&\sqrt{\frac{\pi}{2\left(\ell+\frac{1}{2}\right)}}\left(\frac{{\rm d}z}{{\rm d} r}\right)_{r_{\nu}}\int {\rm d}z'
\bar p(z'|z_{\nu}) U^{\rm Limber}_{\ell}(r(z'),k'),\nn
\ea
where $r_{\nu}=(\ell+\frac{1}{2})/k_1$ and $z_{\nu}$ are the
photometric redshifts. The extra factor of
$\left(\frac{{\rm d}z}{{\rm d} r}\right)_{r_{\nu}}$
comes from the fact that the Bessel integral in equation (\ref{Gcont})
is over $z$ not $r$. 

By substituting the Bessel function approximation directly into the shear
estimator, equation (1), it can be seen that the Limber approximation 
leads to an equivalence between the 3D $k$ and $\ell$ values and
the tomographic $r$ and $\ell$ values. We will discuss the Limber approximation 
further in Section \ref{Tomography from 3D Cosmic Shear}.

\subsection{Convergence of the 3D Cosmic Shear}
\label{Convergence of the 3D Cosmic Shear}

\begin{figure*}
 \centerline{
 \includegraphics[angle=0,width=1.6\columnwidth]{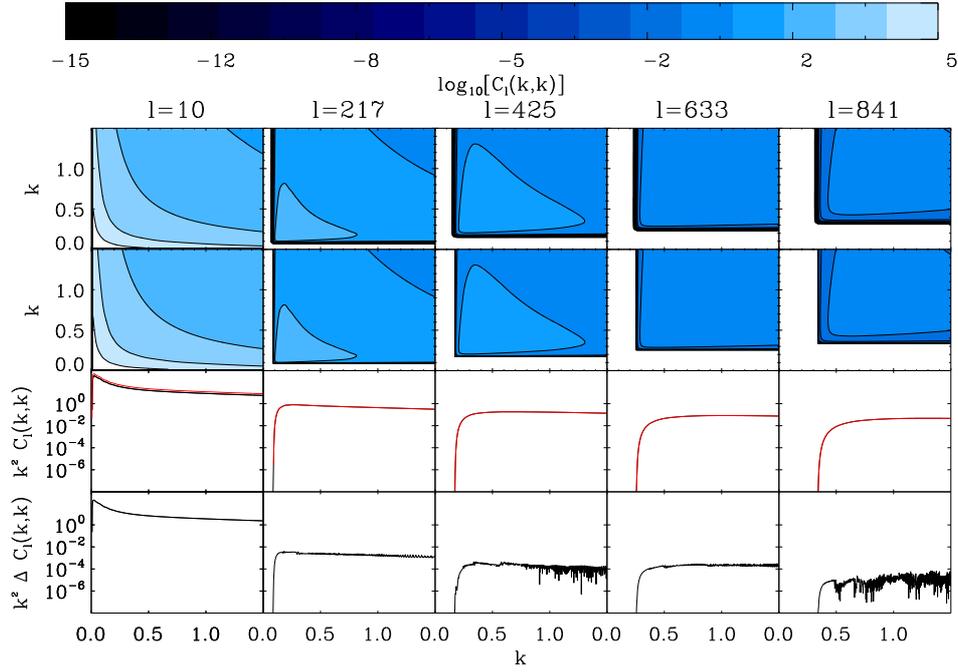}}
 \caption{The 3D cosmic shear power spectra for the full 3D cosmic
   shear calculation (upper panels), and using the Limber
   approximation 
   (second panels). The colour scale is
   shown in the bar above. The third set of panels show the diagonal
   $C_{\ell}(k,k)$ of the full 3D power spectra for the full
   calculation (black lines) and for the Limber approximation (red, dark
   grey lines). The bottom panels show the modulus of the  
   difference in the diagonal elements. This is for a Euclid wide survey
   (Refregier et al., 2010).}
 \label{limberplot}
\end{figure*}

In Figure \ref{limberplot} we show the effect of the Limber
approximation on the 3D cosmic shear
power spectrum. We find that the Limber approximation is a remarkably
good approximation to the full calculation for scales $\ell\gs  
100$, with residuals of $\sim 10^{-4}$ over all radial and 
azimuthal scales. 
The break at
$k\sim \ell/r_{\rm max}$ in the power spectra at each $\ell$, caused
by the Bessel function inequality $j_{\ell}(kr \ls \ell)\sim 0$, is
reproduced through the inequality expressed after equation
(\ref{l2}).  For larger scales $\ell < 100$, 
there is a larger effect on the power
spectrum. 

\begin{figure*}
 \centerline{
 \includegraphics[angle=0,width=1.6\columnwidth]{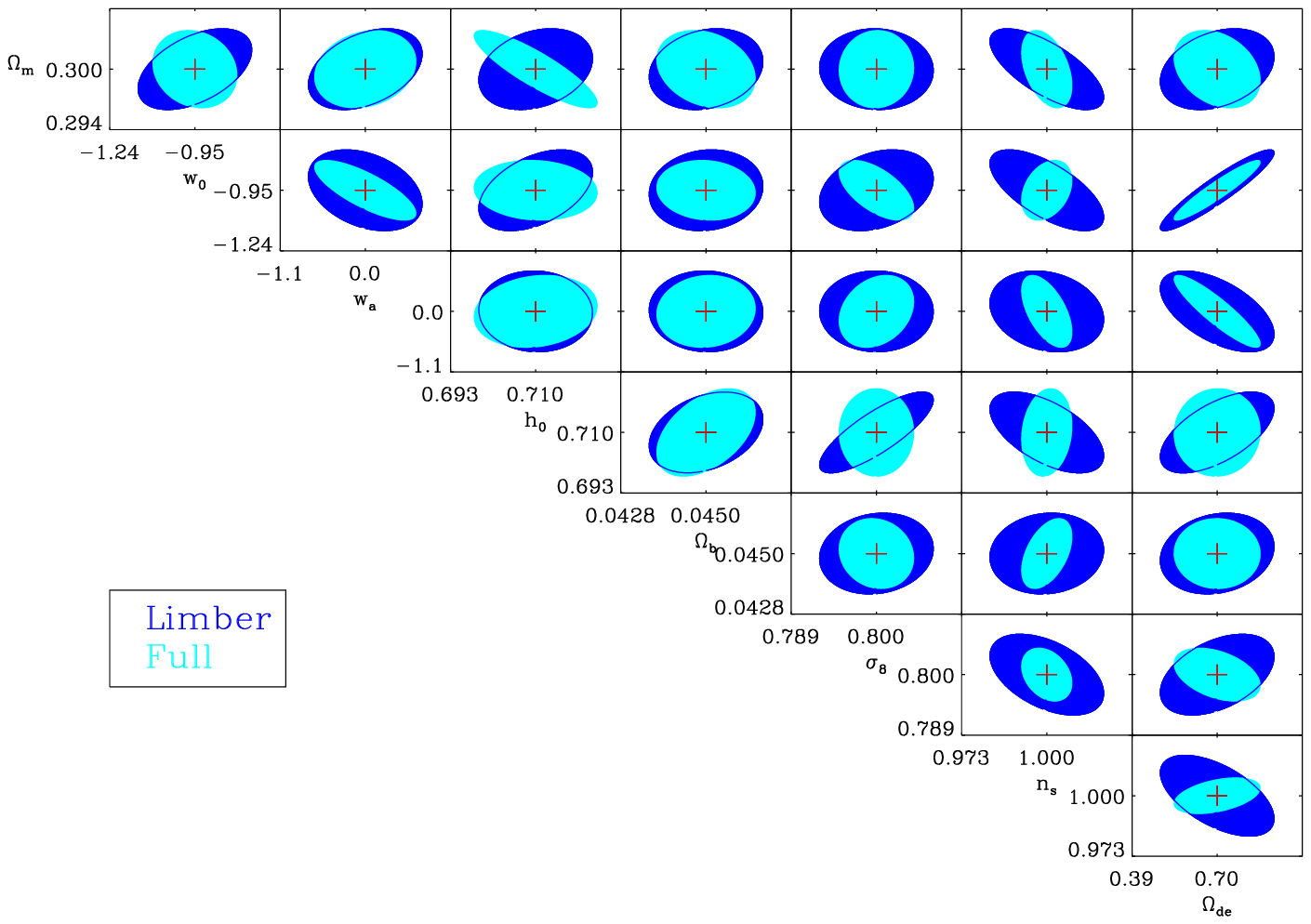}}
 \caption{The Fisher matrix $2$-parameter $1$-$\sigma$ constraints for
   the full 3D cosmic shear calculation and using the Limber
   approximation. Where filled the contours lie behind one another we outline
   the contours with a line. This is for a Euclid wide survey (Refregier et al., 2010).}
 \label{limberfisherplot}
\end{figure*}

In Figure \ref{limberfisherplot} we show the effect of the Limber
approximation on the expected cosmological errors. We find that in
most parameter combinations some
information is inevitably lost, through the largest scales being down
weighted, the increase in errors is between $1$--$30\%$ 
with an average increase of $\sim 10\%$. 

We conclude that the Limber approximation is adequate for forecasting
purposes, but if computer time allows, it is  
preferable to use the full expressions in data analysis.

\section{Tomography from 3D Cosmic Shear}
\label{Tomography from 3D Cosmic Shear}

We now approximate the 3D shear field further and show how 
in the discrete real-space limit the tomographic power spectra can be
reproduced. There have been some implicit references to this derivation,
for example in Hu (1999), here we will show explicitly how the 3D
shear field is related to the tomographic power spectrum. 

Weak lensing tomography is a flavour of 3D weak lensing in which the
angular and redshift information of each galaxy is used. The practical 
distinction between 3D cosmic shear and tomography is that tomography
divides the redshift range into a series of bins and the 2D shear transform in each bin is constructed. The auto (in a
single bin) and cross (between bins) power spectra are used to
constrain cosmological parameters. 

Given the expressions in equations (\ref{l2}) to (\ref{l4}) we can now
derive the weak lensing tomographic power spectra directly from the 3D shear
field. In Appendix B we show how the 3D cosmic shear power spectrum
using the Limber approximation can be written as
\ba
\label{Limb}
&&C^{3D,\rm
  Limber}_{\ell}(k_1,k_2)=\nn
&&\frac{9\Omega^2_mH^4}{4c^2}
\int {\rm d}{r}\frac{P(\ell/r;{r})}{a^2(r)}
\frac{{\mathcal W}(r_1,r){\mathcal W}(r_2,r)}{r^2}
\ea
where we define a weight factor as
\ba
\label{www1}
{\mathcal W}(r_i,r)=[n(r_i)r_i^2]r\int {\rm d}r'\bar p(r')\frac{r-r'}{r'},
\ea
where $r_i = \ell/k_i$ and $p(r[z])=p(r[z]|r[z_p])$ is the redshift probability
distribution, equivalent to
$\bar p(z|z_p)$ in equation (\ref{Gcont}). We condense the notation here and in
Appendix B to match the literature for the tomographic case.
This is still a full 3D estimator where $r_1=r(z_1)$ and $r_2=r(z_2)$ 
can take any value, and in practice would be a sum over all galaxy
pairs. This is a key result of this article, using only two integrals
a full 3D shear power spectrum can be computed from the 3D matter
power spectrum (as simple as the
standard tomographic approximation). 

Inspection of the previous equations shows that they are the usual
expressions for the (auto- or cross-) power spectrum of tomography,
from which we see that, under the Limber approximation, tomography
samples discrete sets of physical wavenumbers, in an $\ell$-dependent
way: for shells at distances $r_i$, $k = \ell/r_i$.

In summary to convert from 3D cosmic shear to weak lensing
tomography we see that the following three steps must be taken
\begin{itemize} 
\item The {\bf Limber approximation} must be applied to the full 3D shear estimator.\\
\item {\bf Fourier transform} the kernel integration from harmonic space to real space.\\
\item {\bf Discretisation} of physical modes through $k=\ell/r_i$.
\end{itemize}
The second step is benign in that no information should be lost,
however the first and third steps do result in information loss (see
Section \ref{Convergence of the 3D Cosmic Shear} and \ref{Comparison
  to 3D shear}).  Interestingly 
for a specific redshift bin ($r$) and a specific azimuthal $\ell$-mode 
the tomographic approximation only probes a single physical
$k$-mode $k=\ell/r$ from the full 3D shear field; in contrast in 3D
cosmic shear we have control over 
the $k$ and $\ell$ modes over the whole redshift range.   

Clearly by fixing the distances of the tomographic binning, we lose
some flexibility over the physical wavenumbers probed,  
so there is a risk that either not all useful modes are included
(increasing statistical errors), or that, for the nearby shells, the
physical wavenumber range sampled extends to too high a value of $k$,
where theoretical uncertainties become a potential source of
systematic error.    None of this is a fundamental problem for
tomography; it simply requires that the $\ell$ range chosen should be
redshift-dependent -- increasing $\ell_{\rm max}=r[z]k_{\rm max}$ 
for the distant shells, and reducing it for nearby shells. 

In a similar manner to Section \ref{Photometric 3D Shear Estimator} we
can also keep the summation over galaxies explicit in the derivation
of the tomographic power spectrum. In this case we have a very similar
expression to equation (\ref{www1}), with the kernel functions are
now replaced by
\ba 
\label{t3}
{\mathcal W}_{\rm individual}(r,r_i)&=&
\sum_{h}r\int {\rm d}r'p_h(r'|r^h)\frac{r-r'}{r'}\nn
&=&r\int {\rm d}r'\left[\sum_{h} p_h(r'|r^h)\right]F_k(r',r), 
\ea 
where we have a summed over the set of galaxies $h=\{g : r_i-\Delta r\leq r
  \leq r_i + \Delta r\}$ within a bin defined by a width $2\Delta
  r$ centred on $r_i$. $p_h(r)$ is the redshift probability distribution for an
  individual galaxy within that bin.

\subsection{Convergence of 3D shear} 
\label{Comparison to 3D shear}

Here we investigate the predicted cosmological error constraints
from  3D cosmic shear as more modes are included. We assume a
survey configuration as given in Section 
\ref{Photometric 3D Shear Estimator}, and restrict the analysis to
the regime $\ell_{\rm max}=5000$ and $k_{\rm max}=1.5$hMpc$^{-1}$ (with 
$N_{\rm modes}=1000$ k-modes evenly distributed over the range), 
we use the full posterior distributions from Bordoloi et al. (2009). 

\begin{figure*}
 \includegraphics[angle=0,clip=,width=\columnwidth]{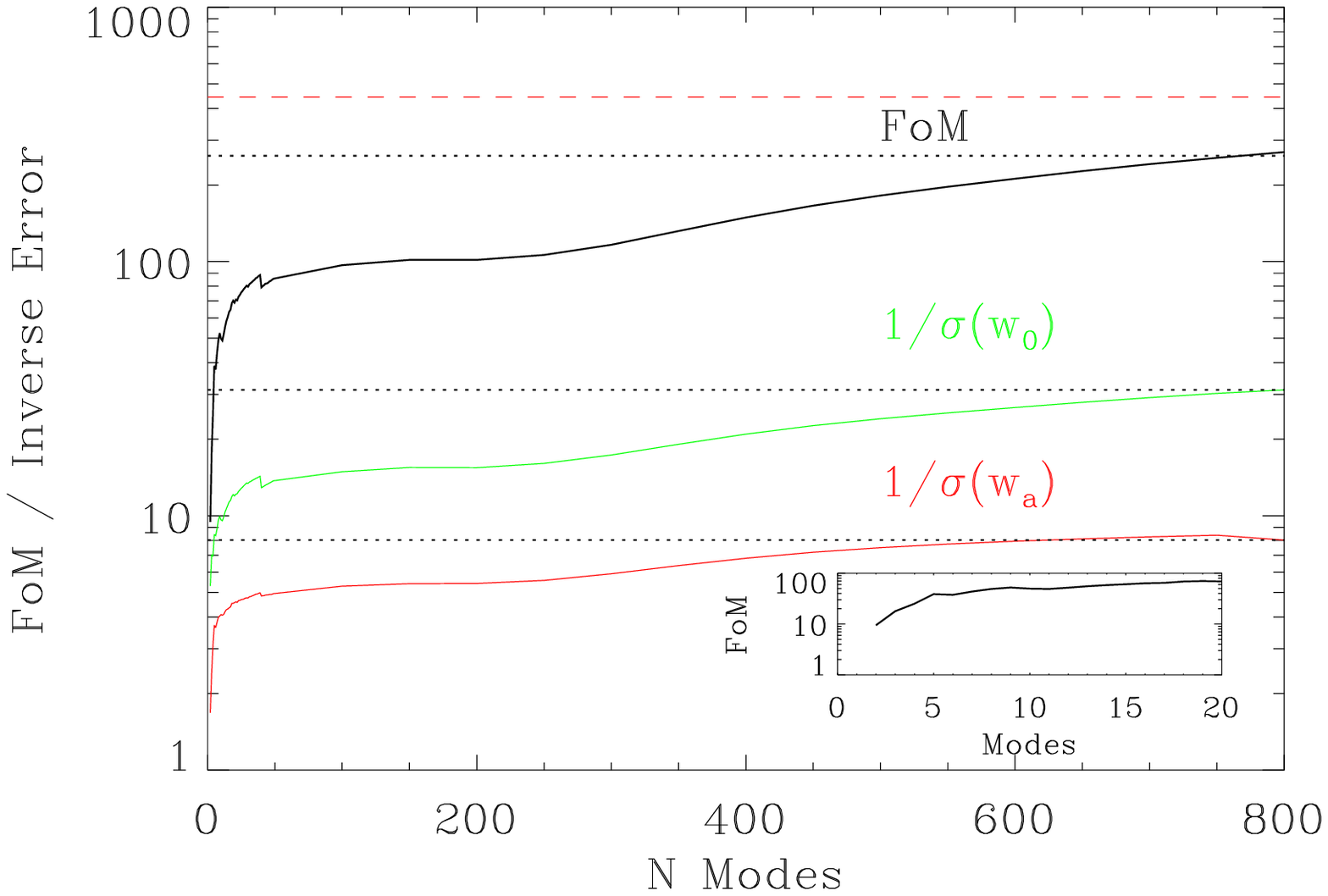}
 \includegraphics[angle=0,clip=,width=\columnwidth]{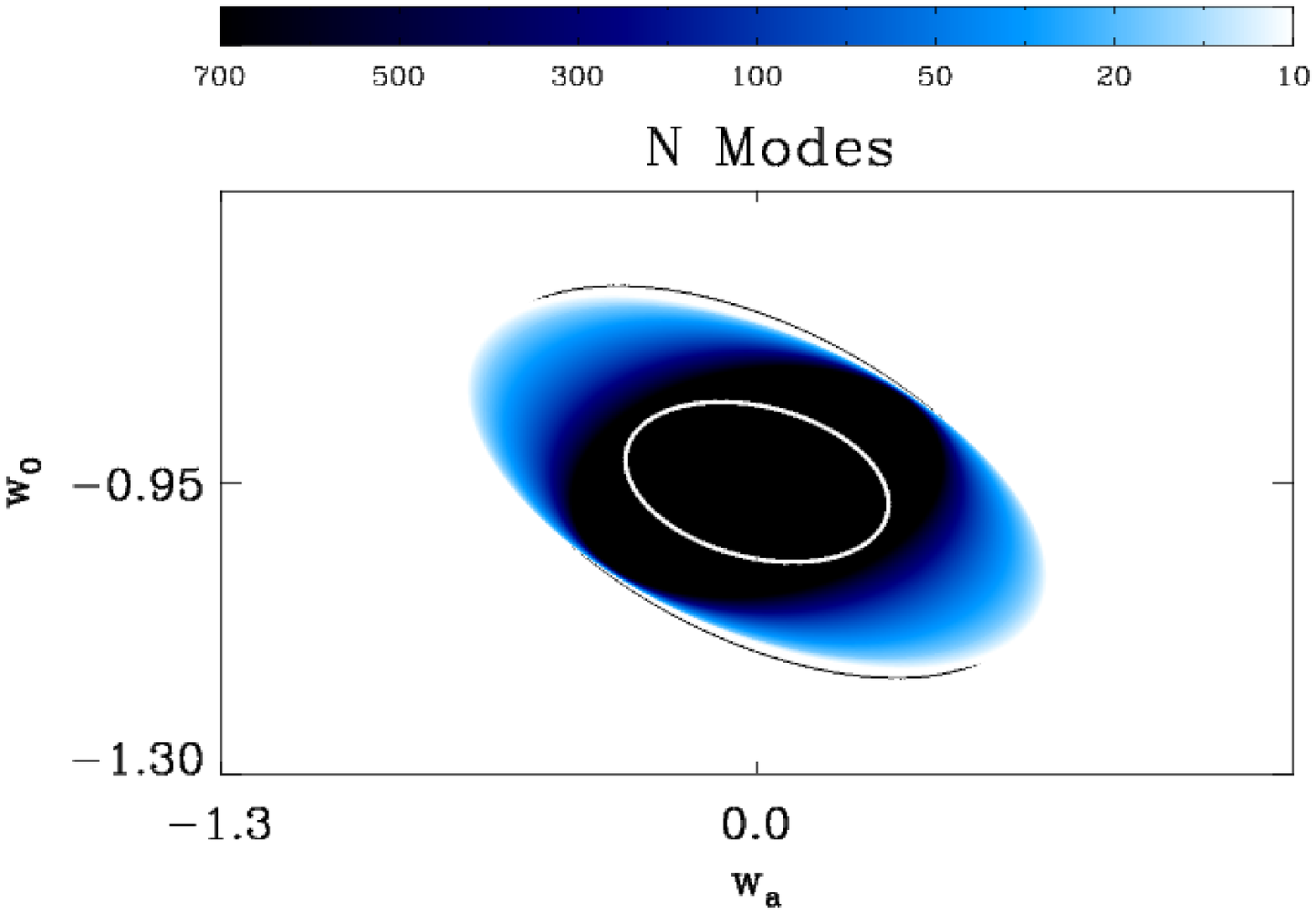}
 \caption{Mode convergence of 3D cosmic shear. For the 3D
 calculation we use the Limber approximation, equation
 (\ref{Limb}). The left panel shows how the inverse errors ($1/\sigma$) 
 for $w_0$ (middle green line), $w_a$ (lower red line) 
 as well as the dark energy Figure of Merit (upper black line) (Albrecht et
 al., 2006) vary with the number of $k$ modes, for  for $\sigma_z(z)/(1+z)=0.03$ . 
 The  dotted horizontal lines show the 3D cosmic shear asymptotic predictions using
 the Limber approximation. The upper dashed (red) line shows the 3D cosmic
 shear prediction using the full (non-Limber) approximation. 
 The inset shows the change in FoM over the range of $2$
 to $20$ modes. The right panel shows the predicted $1$-$\sigma$
 two-parameter 
 constraints in the ($w_0$,$w_a$) plane, the colour
 represents the number of k-mode bins that we define in the
 colour bar, the central white ellipse shows the Limber 
 approximation 3D cosmic shear
 constraints. This is for a Euclid wide survey
 (Refregier et al., 2010).  We use $\ell_{\rm max}=5000$ and
 $k_{\rm max}=1.5$hMpc$^{-1}$.}
 \label{tomoplot1}
\end{figure*}

Figure \ref{tomoplot1} shows how the Fisher matrix errors vary, for $w_0$
and $w_a$ as well as the dark energy Figure of Merit 
(FoM$=1/\sqrt{F^{-1}_{w_0w_0}F^{-1}_{w_aw_a}-(F^{-1}_{w_0w_a})^2}$;
Albrecht et al., 2006),  as a function of the number of $k$ modes sampled. We
also show the 3D cosmic shear constraints using the Limber
approximation. 
In Figure \ref{tomoplot1} we also show how the constraints in the
($w_0$,$w_a$) plane change as the number of modes is increased.
We find that the cosmological constraints do not converge to the 3D
(Limber) limit until the number of modes is 
$\gs 800$ for low-$\ell$ modes. 

We can understand the convergence if we consider that a separation of
modes in $k$-space  
$\Delta k$ for a given $\ell$ corresponds to physical seperations
$\Delta r$ in the following way  
\be 
\label{dk}
\Delta k\simeq \frac{k_{\rm max}}{N_{\rm modes}}\simeq \ell\frac{\Delta r}{r_1r_2}
\ee
where $\Delta r=r_1-r_2$ and $r_i=\ell/k_i$. We can use this as an approximate 
model to investigate the convergence properties. 

\begin{figure*}
 \includegraphics[angle=0,clip=,width=\columnwidth]{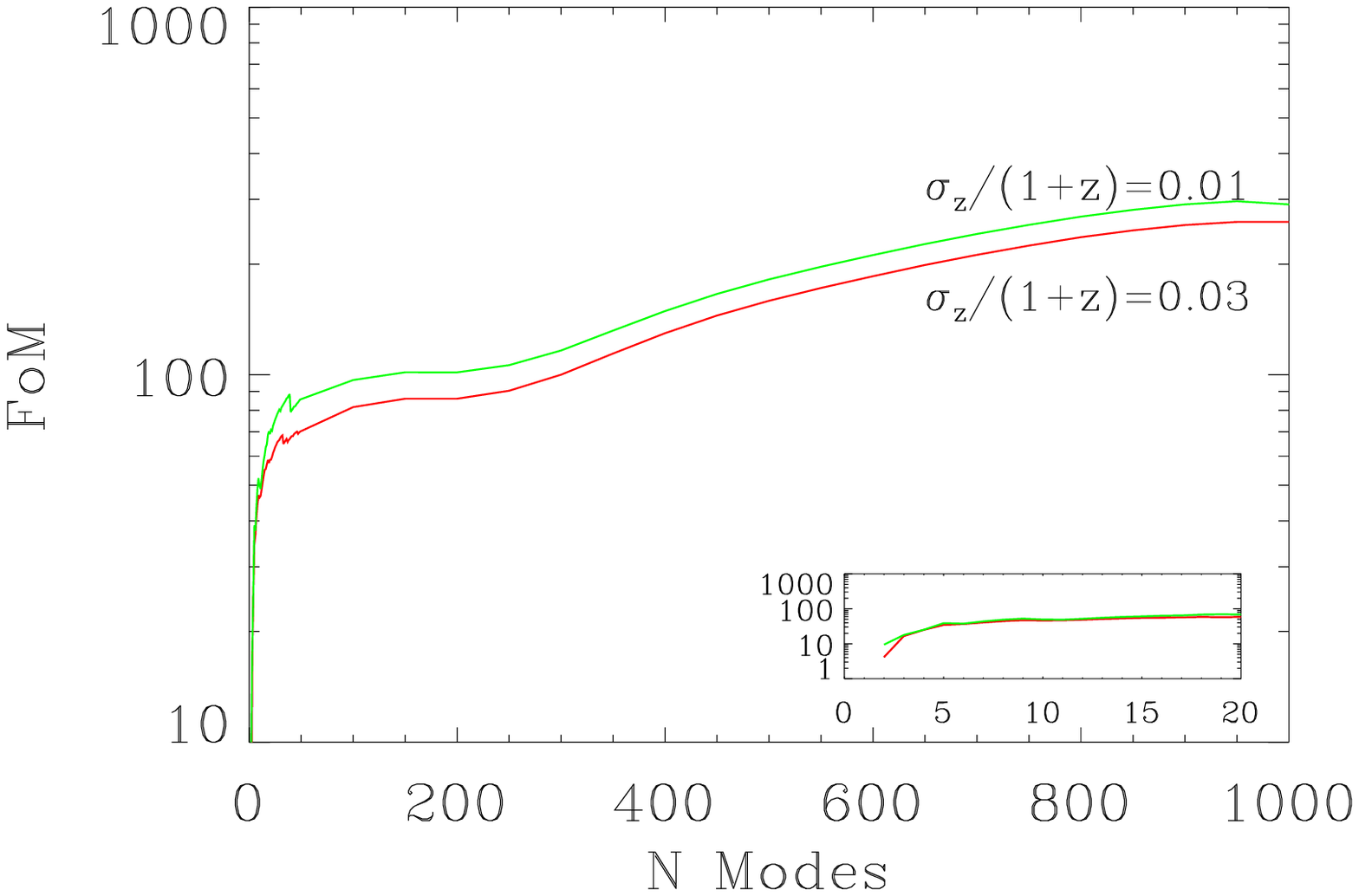}
 \includegraphics[angle=0,clip=,width=\columnwidth]{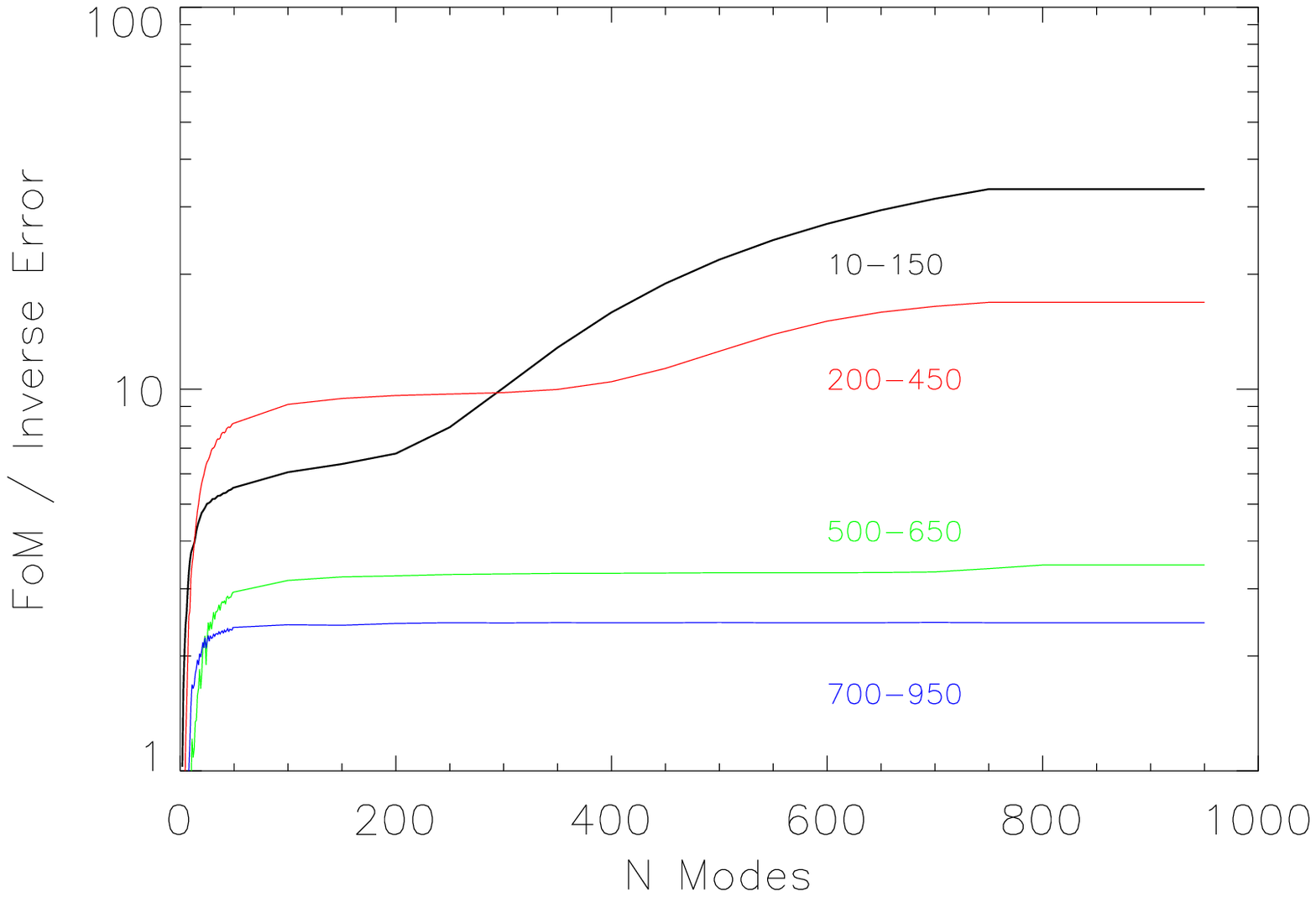}
 \caption{Left : We show the
   dark energy Figure of Merit (Albrecht et
   al., 2006) as a function of the number of $k$ modes analysed. 
   The lower/upper (red,green) 
   lines are for $\sigma_z(z)/(1+z)=0.03$ and $0.01$ 
   The inset shows the change in FoM over the range of $2$
   to $20$ bins for $\sigma_z(z)/(1+z)=0.03$ (red,lower) and
   $\sigma_z(z)/(1+z)=0.01$ (green,upper). This is for a Euclid wide survey
   (Refregier et al., 2010). We use a $\ell_{\rm max}=5000$ and
   $k_{\rm max}=1.5$hMpc$^{-1}$.
   Right : For $\sigma_z(z)=0.03$ we show how the FoM changes for particular 
   $\ell$ ranges (labelled near each line). Note that the in the full 
   calculation, Figure \ref{tomoplot1}, we use a range $\ell=10$--$5000$, 
   and that the largest contribution to the Fisher matrix 
   $F_{ij}=\int{\rm d}\ell\ell F_{\ell,ij}$ (equation \ref{fish}) is in the 
   range $\ell\approx 1000$ (Heavens et al., 2006).}
 \label{tomoplot2}
\end{figure*}

We expect that convergence will occur in two 
regimes i) as radial modes enter the survey volume, 
ii) as radial modes become correlated due to the photometric/$n(z)$ 
smoothing scale.
To illustrate this we use equation (\ref{dk}). The depth of the
fiducial survey in this article  
is $\Delta r_{\rm survey}\approx 3000$h$^{-1}$Mpc (where $r_1\sim 3000$ 
and $r_2\sim 0$) such that $N_{\rm modes}\approx 4500/\ell$. For $\ell=150$ 
this should result in convergence at $N_{\rm modes, survey}\approx 30$. 
The photometric/$n(z)$ smoothing scale 
typically occurs at $\Delta r_{\rm photoz}\approx 150$h$^{-1}$Mpc in comoving 
coordinates in this case we see that $N_{\rm modes,photoz}\approx 95,000/\ell$ 
from equation (\ref{dk}), for $\ell=150$ this should create convergence at 
$N_{\rm modes}\approx 600$. In Figure \ref{tomoplot2} we see that for all $\ell$
ranges the convergence occurs at the photometric/$n(z)$ smoothing scale. For 
high-$\ell$ modes the survey convergence is not visible since 
$N_{\rm modes,survey}\ls 10$. For 
low-$\ell$ modes the difference in the convergence regimes 
($30$ and $600$ k-bins) is clearly visible. 

Note that this simple illustration is 
approximate since the radial smoothing scale is also a function of the 
lensing kernel. In Figure \ref{tomoplot2}
we also show the 
the FoM convergence for
$\sigma_z(z)/(1+z)=0.01$ and $0.03$. 
In both cases the
the constraints improve sharply over the range $1$--$20$ modes. This suggests that
below $\sigma_z(z)/(1+z)\approx 0.03$ the 
radial smoothing scale is not significantly reduced, due to the 
lensing kernel. The change in 
$\sigma_z(z)$ has a sub-dominant effect on the FoM especially at low 
numbers of modes, which is in agreement with tomographic studies (e.g. 
Bridle \& King, 2007) and in the 3D limit with Heavens et al. (2006).

We find that 
under the same survey assumptions, and a restriction to
modes $\ell<5000$, $k<1.5$hMpc$^{-1}$ that the
3D cosmic shear predictions agree very well
with publicly available tomographic code (for example {\tt iCosmo};
Refregier et al., 2008) and other tomographic studies 
(e.g. Amara \& Refregier, 2007; Bernstein, 2009). 
We note however that typical tomographic studies use highly 
non-linear modes, $\ell\sim 10^4$ and $k\gg 1.5$hMpc$^{-1}$,  
and consequently report tighter constraints.

We emphasise here that our results are consistent and complimentary 
with tomographic studies. We find a convergence at the photometric
smoothing scale. This is consistent with tomographic studies that
find convergence at around $20$ redshift 
bins which is also consistent with the
photometric smoothing scale where 
$\Delta r \approx r_{\rm survey}/r_{\rm photo} \sim 3000/150 = 20$. We
recommend that in any analysis both tomography and 3D cosmic shear
should be peformed
\begin{itemize}
\item
Tomography allows one to explicitly scrutinise the redshift dependence
of the signal.\\ 
\item 
3D cosmic shear allows for an explicit radial and azimuthal scale
dependence to be investigated.\\ 
\end{itemize}  
The redshift and scale dependence should be explicitly investigated by 
both methods respectively. 

\section{Conclusion}
\label{Conclusion}

In this article we have developed the 3D cosmic shear technique in a
number of ways. We have simplified the analysis using the Limber
approximation,  
we have shown how to include individual posterior redshift probability
distributions from photo-zs in the analysis.  
The technique presented here removes the need for any
intermediate $p(z|z_p)$ estimation, allowing the estimator itself to
become a direct function of the individual galaxy redshift probabilities.  

Finally, we have
clarified the relationship between 3D cosmic shear and tomography, and
demonstrated that tomography essentially provides an $\ell$- and
shell-dependent sampling of physical wavenumber modes.    

To study individual redshift errors, we used a mock catalogue of photometric redshift 
probabilities $p(z)$ from Bordoloi et al. (2009). 
We find that the cosmological errors can be mis-estimated by
$10$--$50\%$ 
by using an average $\bar p(z|z_p)$ in the estimator, but that biases
on the cosmological  
parameters are negligible. In order to calculate the biases we have
generalised the bias formalism  
introduced in Knox et al. (1998) to the case where the cosmological  
signal is in the covariance not the mean. 

We have not investigated systematic effects that may be present in
photometric redshift estimates.  
However we note that the outlier mitigation techniques and likelihood
calibration methodology of  
Bordoloi et al. (2009) is relevant and should be used in conjunction
with the techniques presented  
here, and leave this for future investigation. 
In addition we refer to Kitching et al. (2008) (systematic
effects on 3D cosmic shear) where a factor of $2$ degradation in FoM
is expected as a result of the primary weak lensing systematics. 

Using the extended Limber approximation of LoVerde \& Afshordi (2008) 
we have found a much simpler expression for the 
3D shear field in the high-$\ell$ regime. For angular
wavenumbers $\ell \gs 100$ the Limber approximation
is essentially exact with a difference between the Limber and
full calculation power spectra of $\ls 10^{-4}$; a fractional difference of 
$\ls 10^{-2}$.

Finally we explicitly derive the weak lensing tomographic power
spectrum from the 3D cosmic shear field. We find that tomography
essentially probes discrete sets of physical wavenumbers, 
which depend both on $\ell$ and on the distance to each individual
tomographic bin.  

The choice of tomographic bin positions 
constrains the 3D modes which can be analysed. The modes which should
be probed have to satisfy several  
constraints: first,  the maximum physical wavenumber should not be too
high, to avoid the uncertain highly non-linear regime; secondly, as
many modes as possible should be included, to reduce errors; thirdly
the effective distance probed by adjacent modes should not be much
less than that set by the photo-z error.  The last of these is most
easily effected with tomography, and the first two within 3D cosmic
shear. 3D cosmic shear automatically deals (in principle) with the
last constraint, through correlation of modes, but too finely-spaced
modes may lead to near-singular covariance matrices. To deal with the
first two constraints in tomography, the $\ell$ range chosen should be
varied with shell.  

The advantages of 3D cosmic shear are in being able to
ensure that appropriate physical modes are analysed, in an integrated
analysis of the entire sample. On the other hand tomography allows
for a straightforward investigation of the redshift-dependence of the
shear signal. Consequently we conclude that both approaches have
their advantages and it is sensible to do both. 


\noindent{\em Acknowledgements:} We especially thank Rongmon Bordoloi 
for use of the photometric redshift estimates from Bordoloi et al.,
(2009) and Filipe Abdalla for use of redshift estimates from Abdalla et al.,
(2007) in an earlier draft. We thank Adam Amara for a careful reading
of an early draft and an anonymous referee for many helpful comments. 
We also thank Andy Taylor and Fergus Simpson for many useful discussions. 
TDK is supported by a STFC Rolling Grant RA0888.



\onecolumn
\section*{Appendix A: 3D Cosmic Shear}

In this appendix we generalize the results of Heavens et al., (2006) and
Kitching et al. (2007). We compute the covariance of the harmonic
shear coefficients utilizing the individual posterior galaxy redshift
probability distributions, rather than simplified sample properties 
(i.e. an averaged $p(z|z_p)$). 

Starting with equation (\ref{1}) we can begin to derive a theoretical
expression for the expected coefficient values by linking the shear
values to the Newtonian potential $\Phi$ via  
$\gamma=\frac{1}{2}{\tilde\partial}{\tilde\partial} \phi$ where $\phi$ is the lensing
potential (${\tilde\partial}=\partial_i+i\partial_j$) we can write a theoretical estimator like 
\be 
\label{est}
\hat\Gamma_{ij}(k,\bell)=\left(\sum_g\sqrt{\frac{2}{\pi}}\phi
j_{\ell}(k r^g_0){\rm e}^{-i\bell.\thetab^g}W(r^g_0)\right)_{ij},
\ee
where $i,j=\{1,2\}$ and 
$\hat\gamma_1(k,\bell)=\frac{1}{2}\left(\hat\Gamma_{11}-\hat\Gamma_{22}\right)$
and $\hat\gamma_2(k,\bell)=\hat\Gamma_{12}=\hat\Gamma_{21}$. 

The covariance of equation (\ref{est}) can be written as 
\ba
\label{est2}
\langle \hat\Gamma(k_1,\bell_1) \hat\Gamma^*(k_2,\bell_2)\rangle=
\sum_g \sum_h \left(\frac{2}{\pi}\right)j_{\ell_1}(k_1
r^g_0)j_{\ell_2}(k_2 r^h_0)
\langle \phi\phi^*\rangle{\rm
  e}^{-i\bell_1.\thetab^g}{\rm
  e}^{+i\bell_2.\thetab^h}W(r^g_0)W(r^h_0)
\ea
where the covariance of the lensing potential can be written in terms
of the Newtonian potential $\Phi$ (note we assume General Relativity and no anisotropic stress, so the Newtonian and curvature potentials are
identical):
\ba
\label{est3}
\langle \phi(r^g_0,\thetab^g)\phi^*(r^h_0,\thetab^h)\rangle=
\frac{4}{c^2}
\int_0^{r^g}{\rm d}r'\int_0^{r^h}{\rm d}r''F_K(r^g,r')F_K(r^h,r'')
\langle \Phi(r',\thetab^g)\Phi^*(r'',\thetab^h)\rangle
\ea
where $F_k(r,r')\equiv S_k(r-r')/[S_k(r)S_k(r')]$ ($S_k=\sinh(k),k,\sin(k)$
for open, flat or closed geometries so $F_k(r,r')=(r-r')/rr'$ for a
flat universe). The real space Newtonian potential can also be written in terms
of its spherical harmonic transform so that   
\be
\label{est4}
\Phi(r,\thetab^g)=
\left(\frac{2}{\pi}\right)^{1/2}\int \frac{{\rm d^2}\bell}{(2\pi)^2}
\int {\rm d}k k j_{\ell}(kr^g)\Phi(k,\bell; t){\rm e}^{i\bell.\thetab^g},
\ee
where $t$ expresses the time-dependence. From now on we will identify this time dependence by an $r$ dependence. 
Using Poisson's equation we can write the Newtonian potential's
covariance (diagonal in $\ell$ because of isotropy) 
in terms of the matter power spectrum $P(k; r)$ as  
\ba 
\label{est5}
\langle \Phi(r',\thetab^g)\Phi^*(r'',\thetab^h)\rangle=
\left(\frac{2}{\pi}\right)A^2\int \frac{{\rm d}k}{k^2}
\frac{j_{\ell}(kr')j_{\ell}(kr'')}{a(r')a(r'')}
\sqrt{P(k; r')P(k; r'')}{\rm e}^{-{\rm i}\bell\dot(\thetab^h-\thetab^g)}
\ea
where $A=3\Omega_mH^2/2$. Note that, as shown in Castro et al. (2003)
we take advantage of an algebraic convenience by using the geometric
mean of the matter power spectra, this is well justified since at
large separations (where the approximation may break down) any pair
correlations are down weighted by the form of the Bessel functions. 

This is fed into equations (\ref{est}) then
(\ref{est2}) so that we have an expression for the expected covariance
of the 3D cosmic shear estimator 
\ba 
\label{est6}
\langle \hat\Gamma_{ij}(k_1,\bell)
\hat\Gamma^*_{ij}(k_2,\bell)\rangle=
\Delta\Omega\ell^2_i\ell^2_j\left(\frac{2}{\pi}\right)^2\left(\frac{1}{c^2}\right)A^2
\sum_g\sum_h j_{\ell}(k_1 r^g_0)j_{\ell}(k_2 r^h_0)W(r^g_0)W(r^h_0)\nn
\int_0^{r^g_0}{\rm d}r'\int_0^{r^h_0}{\rm d}r''
F_K(r^g,r')F_K(r^h,r'')
\int \frac{{\rm
    d}k'}{k'^2}\frac{1}{a(r')a(r'')} 
j_{\ell}(k'r')j_{\ell}(k'r'')\sqrt{P(k';r')P(k';r'')},
\ea
the prefactor $\Delta\Omega$, the angular size or area of the survey, 
comes from final integrations over angle $\thetab$ (see Kitching et al., 2007 for
more information). 

We use this final expression in Section \ref{Photometric 3D Shear
  Estimator}.

\newpage

\section*{Appendix B: Tomography from 3D Cosmic Shear}

Here we derive the tomographic shear power spectra from the 3D cosmic
shear covariance (equations \ref{CGU}, \ref{Gcont}), assuming a global
$\bar p(z|z_p)$.

We start with the full 3D cosmic shear power spectrum as a function of 
azimuthal $\ell$ and two radial $k$ wavenumbers. This is an expansion
of the equations (\ref{CGU}) and (\ref{Gcont}):
\ba 
C^{3D}_{\ell}(k_1,k_2)={\mathcal A}^2
\int {\rm d}r_gr_g^2 n(r_g)j_{\ell}(k_1r_g)
\int {\rm d}r_hr_h^2 n(r_h)j_{\ell}(k_2r_h)
\int {\rm d}r'\bar p(r'|r_g)\int {\rm d}r''\bar p(r''|r_h)\nn
\int {\rm d}{\tilde r}'\int {\rm d}{\tilde r}''
\frac{F_K(r',{\tilde r}')}{a(\tilde r)'}\frac{F_K(r'',{\tilde r}'')}{a(\tilde r'')}\int\frac{{\rm d}k'}{k'^2}
j_{\ell}(k'{\tilde r'})j_{\ell}(k'{\tilde r}'')P^{1/2}(k'; {\tilde r}')P^{1/2}(k'; {\tilde r}'')
\ea
where we see that four Bessel functions enter the expression. The
prefactor ${\mathcal A}^2$ is given by
${\mathcal A}^2=\ell^4\frac{4}{\pi^2c^2}9\Omega^2_mH^4/4$. 
The $p(r[z])=p(r[z]|r[z_p])$ is the redshift probability
distribution, equivalent to
$p(z|z_p)$ in equation (\ref{Gcont}), and we condense the notation here
to match the literature for the tomographic case.

We now replace each Bessel function with its Limber approximated form
(see equation \ref{l1}) and find that we can rewrite the 3D power
spectrum  
\ba
C^{3D,\rm
  Limber}_{\ell}(k_1,k_2)={\mathcal A}^2\left(\frac{1}{\ell}\right)^2\left(\frac{\pi}{2\ell}\right)^2
\int {\rm d}{\tilde r}{\tilde r}^2 \frac{P(\ell/{\tilde r};{\tilde
    r})}{a^2({\tilde r})}
n(r_1)r_1^2n(r_2)r_2^2\int {\rm d}r'\bar p(r')F_K(r',{\tilde r})\int {\rm
  d}r''\bar p(r'')F_K(r'',{\tilde r})
\ea
where we have introduced two radial distances $r_1$ and $r_2$ that are
defined through the Limber approximation as $r_1=\ell/k_1$ and
$r_2=\ell/k_2$. We can rewrite this in a more familiar form 
\ba
C^{3D,\rm
  Limber}_{\ell}(k_1,k_2)=\frac{9\Omega^2_mH^4}{4c^2}
\int {\rm d}{r}\frac{P(\ell/r;{r})}{a^2(r)}
\frac{{\mathcal W}(r_1,r){\mathcal W}(r_2,r)}{r^2}
\ea
where we define a weight factor as
\ba
{\mathcal W}(r_i,r)=[n(r_i)r_i^2]r\int {\rm d}r'\bar p(r'|r_i)\frac{r-r'}{r'}.
\ea
We have written the geometric term in the flat geometry case
for clarity $F_{K=0}(r,r')=(r-r')/rr'$, and the $r^2$ term cancels;
to match convention we add a factor of $r$ to the weight factor that
must be cancelled by a $r^2$ denominator.  

We note here that this expression is still a full 3D estimator in the
Limber approximation. We use this expression in Section 
\ref{Tomography from 3D Cosmic Shear} to
derive the tomographic power spectra and relate this to the 3D cosmic
shear power spectra. 

\section*{Appendix C: Bias for Signal in the Covariance}

The bias formalism presented in Knox et al. (1998), Amara \& Refregier
(2007) and Taylor et al. (2007) assumes that the parameter dependency
comes through the 
mean of the signal. Here we show how the bias formalism can be
extended to the case where the signal is in the covariance. 

We start with a general expansion, in $\Delta\btheta$, of the (log)
likelihood about the maximum likelihood position $\btheta_0+\Delta\btheta$
\be 
{\mathcal L}(\btheta_0+\Delta\btheta)={\mathcal
  L}(\btheta_0)+\Delta\theta_i\partial_i{\mathcal L}(\btheta_0)+\dots 
\ee
where $\btheta_0$ in this case is the old position of the maximum likelihood
and $\Delta\btheta$ is the bias in parameters to a new maximum
likelihood. By taking the derivative with respect to the parameters we
have
\be
\partial_j{\mathcal L}(\btheta_0+\Delta\btheta)=2\partial_j{\mathcal
  L}(\btheta_0)+\Delta\theta_i\partial_i\partial_j{\mathcal
  L}(\btheta_0)+\dots = 0
\ee
where the log-likelihood is maximised at ${\mathcal
  L}(\btheta_0+\Delta\btheta)$. By taking the expectation value of
this and rearranging we find that the bias $b_i$ in the $i^{\rm th}$
parameter can be written as 
\be 
b_i=\Delta\theta_i=F^{-1}_{ij}\langle 2\partial_j{\mathcal
  L}(\btheta_0)\rangle,
\ee
in a completely general way. Note that this makes the assumption that
the Fisher matrix is unaffected to first order (this is a good
approximation for small systematics as shown by Joachimi \& Bridle,
2009). 

By taking the result from Tegmark, Taylor \& Heavens (1997) in the
Gaussian case we can write the first derivative of the log-likelihood as 
\be 
\label{c1}
(2{\mathcal L},_i)_{\ell}={\rm
  Tr}\left[C_{\ell}^{-1}C_{\ell},_i-C_{\ell}^{-1}C_{\ell},_iC_{\ell}^{-1}D+C_{\ell}D,_i\right] 
\ee
as a function of $\ell$, 
where we replace $\partial_i=,_i$. $D$ is the data matrix
$(x-\mu)(x-\mu)^t$. Here and for the remainder of this Appendix we assume an all sky survey such that 
the covariance $C_{\ell}$ is diagonal in $\ell$ (see Heavens et al., 2006; Kitching et al., 2007).
We now assume that the mean is not a function of
parameters and only consider the case where the parameters are in the
covariance $D,_i=0$; in the case that the signal is in the mean and
$C_{\ell},_i=0$ we recover the result of Amara \& Refregier (2007) and
Joachimi \& Bridle (2009).  Note that the quantity in the square brackets is a
$(\ell,k,k')$ 3D matrix, and that the derivative of the log-likelihood is a
function of $\ell$.

To include a systematic we rewrite the covariance
as some true covariance plus a systematic $C=C^t+C^s$. By taking the
expectation value of equation (\ref{c1}) we have
\ba
\label{c2}
\langle 2{\mathcal L},_i\rangle_{\ell}={\rm
  Tr}[(C_{\ell}^t+C_{\ell}^s)^{-1}C_{\ell}^t,_i-
  (C_{\ell}^t+C_{\ell}^s)^{-1}C_{\ell}^t,_i(C_{\ell}^t+C_{\ell}^s)^{-1}\langle
  D\rangle]
\langle 2{\mathcal L},_i\rangle_{\ell}={\rm
  Tr}\left[(C_{\ell}^t+C_{\ell}^s)^{-1}C_{\ell}^t,_i\left(I-(C_{\ell}^t+C_{\ell}^s)^{-1}C_{\ell}^t\right)\right]
\ea
where the expectation of the data is the true covariance $\langle D\rangle=C_{\ell}^t$.
Finally we note that the total log-likelihood is in general a sum over
$\bell=(\ell_x,\ell_y)$ modes ${\mathcal L}=\sum_{\ell_x,\ell_y}{\mathcal L}_{\ell_x,\ell_y}$, so a similar
summation must be performed over the derivatives of the log-likelihood. For the
Fisher matrix and for the bias here we write this summation 
as an integral over the modulus $\ell=|\bell|$, and account
for the density of state in $(\ell_x,\ell_y)$ (see Appendix B in Kitching, Taylor, Heavens, 2007).

We can now write an expression for the bias caused by a systematic
function in the case of the parameters being in the covariance like
\ba
\label{c3}
b_i=F^{-1}_{ij}\big\{g\int{\rm d}\phi_{\ell}\int {\rm d}\ell \ell
{\rm Tr}\left[(C_{\ell}^t+C_{\ell}^s)^{-1}C_{\ell}^t,_j
\left(I-(C_{\ell}^t+C_{\ell}^s)^{-1}C_{\ell}^t\right)\right]\big\}.
\ea
In this final step we have also added an integration over $\ell$-space
where $g$ is the density of states in $\ell$; this is exactly the same
integration that is performed when calculating the Fisher matrix (see
Appendix B in Kitching, Taylor, Heavens, 2007).
If the systematic is very low $|C^s|\ll C^t$ then
$\langle 2{\mathcal L},_i\rangle = 0$ and the bias is effectively zero.

\end{document}